\documentclass{article}

\usepackage{arxiv}

\usepackage[utf8]{inputenc}
\usepackage[T1]{fontenc}
\usepackage{hyperref}
\usepackage{url}
\usepackage{booktabs}
\usepackage{amsfonts}
\usepackage{nicefrac}
\usepackage{microtype}
\usepackage{lipsum}
\usepackage{graphicx}
\usepackage{amsmath}

\usepackage{listings}
 \usepackage{amsmath,amssymb,mathtools}
 \usepackage{algorithm}
 \usepackage{algpseudocode}
 \newcommand{\sysname}{\textsc{pokiSEC}}

\title{pokiSEC: A Multi-Architecture, Containerized Ephemeral Malware Detonation Sandbox}

\author{
\begin{tabular}{c}
Alejandro Avina$^{1}$, Yashas Hariprasad$^{1}$, Naveen Kumar Chaudhary$^{2}$ \\[0.6em]
 $^{1}$Department of Computer Science, California State University, East Bay, USA \\
 $^{2}$National Forensic Sciences University, Gandhinagar, India \\[0.4em]
 \texttt{afernandez71@horizon.csueastbay.edu, yashas.hariprasad@csueastbay.edu} \\
\texttt{naveen.chaudhary@nfsu.ac.in}
\end{tabular}
}

\begin{document}

\maketitle
\begin{abstract}
Dynamic malware analysis requires executing untrusted binaries inside strongly isolated, rapidly resettable environments. In practice, many detonation workflows remain tied to heavyweight hypervisors or dedicated bare-metal labs, limiting portability and automation. This challenge has intensified with the adoption of ARM64 developer hardware (e.g., Apple Silicon), where common open-source sandbox recipes and pre-built environments frequently assume x86\_64 hosts and do not translate cleanly across architectures. 

This paper presents pokiSEC, a lightweight, ephemeral malware detonation sandbox that packages the full virtualization and access stack inside a Docker container. pokiSEC integrates QEMU with hardware acceleration (KVM when available) and exposes a browser-based workflow that supports \emph{bring-your-own} Windows disk images. The key contribution is a \emph{Universal Entrypoint} that performs runtime host-architecture detection and selects validated hypervisor configurations (machine types, acceleration modes, and device profiles), enabling a single container image and codebase to launch Windows guests on both ARM64 and x86\_64 hosts. We validate pokiSEC on Apple Silicon (ARM64) and Ubuntu (AMD64), demonstrating interactive performance suitable for analyst workflows and consistent teardown semantics via ephemeral container lifecycles. The implementation is available at \url{https://github.com/PanLuvme/pokiSEC}.
\end{abstract}


\section{Introduction}
Malware analysis increasingly depends on \emph{dynamic} observation, executing suspicious artifacts and recording what they do in a controlled setting \cite{hariprasad2024securing, wang2023ai}. In contrast to static reverse engineering, detonation-based workflows expose runtime behaviors such as process creation, registry modifications, dropped payloads, privilege escalation attempts, and outbound network activity \cite{miller2022cyber, barham2003xen, vrable2005potemkin, soltesz2007containers}. The central requirement is therefore robust isolation. The analysis environment must be sufficiently separated from the analyst’s host and any surrounding production infrastructure to prevent accidental infection or persistence while still being usable enough to support repeated, iterative inspection \cite{willems2007cwsandbox, yin2007panorama}.

Historically, practitioners have relied on Type-1 hypervisors, dedicated “air-gapped” labs, or carefully curated virtual machine templates for safe detonation \cite{iyengar2025ai, iyengar2025hybrid, iyengar2025privacy, iyengar2025cybersecurity}. These approaches remain effective, but they impose practical overheads that are hard to ignore in modern workflows. They are slow to reset between runs, resource-intensive, and often difficult to automate and integrate with CI/CD-style pipelines or repeatable experimental setups. In parallel, the analyst community has moved toward reproducible tooling and infrastructure-as-code for security research, but malware detonation environments still frequently require bespoke host configuration, specialized virtualization plumbing, and manual environment preparation \cite{iyengar2025cyber, iyengarhb, kumar2023ai}.

A second, newer pressure point is architectural fragmentation \cite{rossow2012prudent}. As ARM64 workstations (notably Apple Silicon) have become mainstream, security teams and researchers increasingly operate mixed fleets where development laptops and some CI runners differ from traditional x86\_64 analysis workstations. Many open-source security tools and pre-built sandbox recipes are optimized for x86 assumptions, and cross-architecture operation is often not a first-class design goal \cite{agache2020firecracker, egele2012survey}. As a consequence, an analyst using an ARM64 laptop may not be able to spin up a standard Windows sandbox without incurring substantial emulation overhead or depending on proprietary virtualization stacks. This mismatch impacts not only performance, but also reproducibility: “the same sandbox” on two different machines may not actually be the same, undermining consistent experimentation and collaborative analysis.

At the same time, containerization has demonstrated compelling benefits for reproducible deployment and packaging. Docker-style workflows make it straightforward to ship complete software stacks, reduce “works on my machine” failures, and simplify onboarding for new contributors. However, malware detonation typically requires a full guest operating system and hardware-assisted virtualization rather than process-level isolation alone. Bridging these requirements motivates an architecture that \emph{packages virtualization itself} inside a container, while preserving near-native performance through hardware acceleration when available \cite{bayer2006dynamic, dinaburg2008ether}.

This paper presents \sysname\, a lightweight, ephemeral malware detonation sandbox designed to run \emph{entirely within a Docker container}. The system encapsulates QEMU/KVM-based virtualization and a web-based ingestion interface into a single, portable image. The key goal is to reduce the operational friction of detonation environments, installation complexity, architecture-specific configuration, and the manual “reset-to-clean-snapshot” loop, while preserving the isolation properties required for safe analysis. In pokiSEC, users can upload a pre-configured Windows disk image (QCOW2) through a web loader and then interact with the booted guest through an in-browser desktop session. The stack is structured so that the host only needs Docker (and virtualization support); the rest of the sandbox dependencies, orchestration logic, and access layer are contained and versioned within the image.

A core design challenge is achieving \emph{multi-architecture portability} without forcing users to maintain separate builds and separate operational playbooks. pokiSEC addresses this via a \emph{Universal Entrypoint} that performs runtime detection of the underlying host architecture and selects the corresponding QEMU binary and acceleration configuration. On ARM64 hosts, the system launches an ARM-targeted QEMU configuration with KVM acceleration, while on x86\_64 hosts it uses the standard x86\_64 hypervisor configuration and KVM enablement. This approach hides architecture-specific QEMU flags behind a consistent execution interface, allowing a single codebase and a single container image to support heterogeneous environments.

Beyond portability, a detonation sandbox must also provide a clean “fresh start” for each run. pokiSEC therefore emphasizes \emph{ephemerality}: the container is intended to be executed with non-persistent semantics (e.g., Docker’s \texttt{--rm}), ensuring that filesystem and system-state changes made during detonation are confined to a temporary copy-on-write layer that is destroyed when the container stops. The resulting workflow reduces the risk of cross-contamination between runs and lowers the operational cost of repeatedly returning to a known-clean baseline, which is essential for iterative malware analysis and controlled experimentation.

The system also prioritizes usability for rapid onboarding and interactive exploration. Rather than requiring CLI-only workflows and manual network plumbing, pokiSEC introduces a \emph{loader-to-hypervisor handoff} mechanism: a lightweight web loader accepts disk images and then relinquishes control so that the hypervisor session can take over on the same endpoint, exposing the guest desktop through a browser-based VNC interface. This design keeps the user experience simple (“upload then boot”) while retaining the flexibility of QEMU/KVM under the hood.

\textit{Summary of contributions: }This paper introduces a containerized malware detonation environment that packages the hypervisor, orchestration logic, and browser-accessible interface into a single Docker image. It proposes a Universal Entrypoint that dynamically reconfigures the virtualization backend based on runtime architecture detection, enabling a unified sandbox workflow across ARM64 and x86\_64 hosts. Furthermore, it demonstrates an ephemeral execution model for detonation that favors rapid teardown and reliable sanitization between runs. Finally, it validates the approach on representative ARM64 and AMD64 hosts, illustrating that containerized virtualization can support near-native interactive workflows on mixed hardware fleets.

\textit{Organization of the paper:} Section~2 reviews related work in malware sandboxes and virtualization/containerization-based approaches. Section~3 presents the system design and methodology of pokiSEC, including the high-level architecture and key components. Section~4 combines implementation details and the proof-of-concept workflow, describing the Universal Entrypoint, the loader-to-hypervisor handoff, and the ephemeral persistence model. Section~5 reports evaluation results across architectures, and Section~6 concludes with limitations and directions for future work.

\section{Related Work}
Dynamic malware analysis has a long history of leveraging sandboxed execution to observe behaviors in a controlled environment. Early and widely used open-source systems such as Cuckoo Sandbox popularized automated detonation pipelines that execute samples inside instrumented virtual machines and collect behavioral artifacts for downstream analysis. While such systems are powerful, they often require complex host configuration, careful dependency management, and tight coupling to specific virtualization assumptions. These costs become more visible when teams attempt to scale analysis across diverse machines, integrate detonation into automated workflows, or ensure consistent environments across collaborators.

Virtualization technologies such as QEMU and KVM form the foundation of many practical sandbox implementations. QEMU provides portable machine emulation and virtualization capabilities, while KVM enables hardware-assisted acceleration on supported hosts, closing the performance gap between emulation and near-native execution. In practice, however, configuring QEMU/KVM for reliable guest execution is non-trivial: machine types, device models, and acceleration flags differ across architectures, and host-specific “known good” configurations are often encoded in ad hoc scripts. This configuration burden is amplified in multi-architecture settings where the same guest workflow must run on ARM64 and x86\_64 hosts.

Containerization has transformed software deployment by packaging dependencies and standardizing runtime environments. Docker in particular has become a de facto mechanism for reproducible tooling distribution and lightweight isolation. Nonetheless, for malware detonation use cases, containers alone are insufficient because malicious behavior is typically observed inside a full guest OS rather than as a single process. As a result, an emerging design pattern is to \emph{containerize the sandbox stack itself}, bundling orchestration logic, virtualization backends, and access interfaces into a portable image. pokiSEC follows this direction while emphasizing two properties that are frequently under-served in practice: first-class cross-architecture execution, and truly ephemeral teardown semantics to reduce cross-run contamination.

Finally, the move toward Windows-on-ARM environments and the broader ARM64 ecosystem creates a new axis in sandbox engineering. While Windows guest support exists for ARM64 hosts, the operational reality for analysts is that many established recipes and toolchains are written with x86\_64 defaults in mind. Bridging that gap requires not only support for ARM64 guests and hosts, but also tooling that hides the architectural complexity behind a consistent user experience. pokiSEC contributes to this space by treating architecture detection and reconfiguration as a primary design goal rather than an afterthought, and by combining that portability with a web-driven detonation workflow that lowers the barrier to entry for interactive analysis.

\section{System Overview and Cross-Architecture Design Methodology}
\label{sec:system}

\subsection{Design Goals and Operating Assumptions}
\label{sec:goals}
pokiSEC is engineered around four primary goals aligned with practical malware detonation needs and reproducible systems research:
\begin{enumerate}
    \item \textbf{Portability across heterogeneous hosts:} a single distribution artifact should execute on both AMD64 and ARM64 machines without per-host rewiring.
    \item \textbf{Low operational friction:} deployment should require only a container runtime plus standard virtualization support; all other dependencies must be bundled.
    \item \textbf{Ephemeral execution by default:} each detonation run should begin from a known baseline and terminate without retaining host-side state.
    \item \textbf{Interactive usability:} analysts should be able to upload a pre-built Windows disk image and interact with the guest through a browser session.
\end{enumerate}

We assume (i) the host kernel supports hardware virtualization (e.g., VT-x/AMD-V on x86\_64 or EL2 on ARMv8), and (ii) the container runtime is permitted to expose the relevant device interfaces (e.g., \texttt{/dev/kvm}) to the container. When such acceleration is unavailable, QEMU can fall back to software emulation; however, the system is optimized for accelerated execution because interactive malware analysis is latency-sensitive.

\subsection{Proposed Solution: Containerized Virtualization as a First-Class Primitive}
\label{sec:proposed}
We present \textbf{pokiSEC}, a containerized detonation sandbox that abstracts operational complexity by packaging the hypervisor and access stack into a single Docker image \cite{merkel2014}. The runtime combines QEMU \cite{bellard2005} with KVM acceleration \cite{kivity2007} and a web-based loader that supports a \emph{Bring-Your-Own-Image} workflow (e.g., Windows QCOW2 images). This design eliminates host-level dependency management and makes the sandbox portable across machines and teams.

Formally, let the host architecture be
\[
\mathcal{A} = \{\texttt{x86\_64}, \texttt{aarch64}\},
\]
and let virtualization availability be $\mathcal{V}=\{0,1\}$ indicating whether hardware acceleration is accessible (e.g., \texttt{/dev/kvm} present and permitted). pokiSEC selects a launch configuration vector
\[
\mathbf{c} = (q, m, \alpha, \rho, \nu) \in \mathcal{C},
\]
where $q$ is the QEMU binary, $m$ is the machine type, $\alpha$ is the acceleration mode, $\rho$ is the device/network profile, and $\nu$ is the remote display profile. The Universal Entrypoint implements
\begin{equation}
\label{eq:cfg-map}
f:\mathcal{A}\times\mathcal{V}\rightarrow \mathcal{C}, \qquad \mathbf{c}=f(a,v),
\end{equation}
ensuring a single code path yields a validated parameter bundle across architectures.

\subsection{High-Level Architecture and Control-Plane/Data-Plane Separation}
\label{sec:arch}
pokiSEC follows a micro-virtualization architecture: the host runs a container engine, and the container encapsulates both (i) a \emph{control plane} for ingestion/orchestration and (ii) a \emph{data plane} for guest execution and interaction.

\paragraph{Control plane (Loader + Orchestrator):}
A lightweight web loader (Flask) accepts a QCOW2 disk image, performs input validation, and places it into a known path. An entrypoint/orchestrator script performs architecture detection, chooses QEMU/KVM configuration, and manages process lifecycles.

\paragraph{Data plane (Hypervisor + Remote Desktop):}
The data plane launches QEMU/KVM and exposes the guest desktop through an HTML5 VNC client, enabling browser-only access for interactive analysis.

\paragraph{Routing and Loader-to-Hypervisor handoff:}
An HTTP reverse proxy multiplexes a single external endpoint between loader UI (initial state) and guest desktop (active state). We model the workflow as a finite-state machine:
\[
\mathcal{S}=\{\textsf{LOADER},\textsf{VM\_RUNNING},\textsf{TERMINATED}\},
\]
with transitions
\[
\textsf{LOADER}\xrightarrow{e_u}\textsf{VM\_RUNNING}\xrightarrow{e_t}\textsf{TERMINATED},
\]
where $e_u$ denotes a successful upload/validation event and $e_t$ denotes termination. The routing policy is
\[
r:\mathcal{S}\rightarrow \{\textsf{route\_loader}, \textsf{route\_vnc}\},
\]
such that $r(\textsf{LOADER})=\textsf{route\_loader}$ and $r(\textsf{VM\_RUNNING})=\textsf{route\_vnc}$.

\begin{figure}[t]
  \centering
  \includegraphics[width=0.95\linewidth]{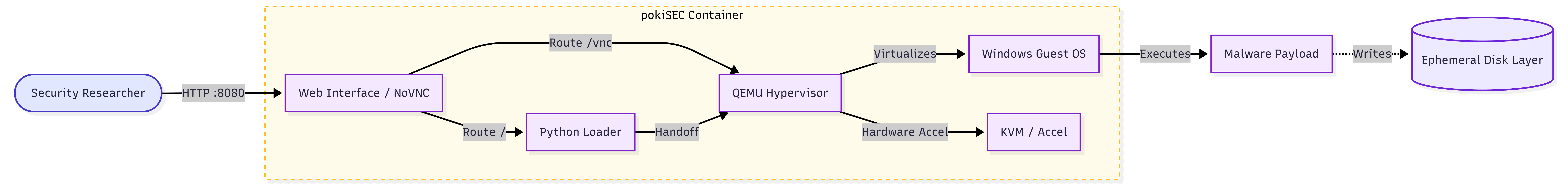}
  \caption{System architecture. A reverse proxy routes the initial session to the loader for image ingestion, then switches to the VM desktop session once QEMU/KVM is launched.}
  \label{fig:arch}
\end{figure}

\subsection{Universal Entrypoint: Architecture-Adaptive Launch Policy}
\label{sec:entrypoint}
At runtime, the entrypoint determines $a\in\mathcal{A}$ and $v\in\mathcal{V}$ and selects acceleration as
\begin{equation}
\label{eq:accel}
\alpha(v)=
\begin{cases}
\texttt{kvm} & \text{if } v=1,\\
\texttt{tcg} & \text{if } v=0.
\end{cases}
\end{equation}
The machine profile $m(a)$ is chosen from a small validated set known to boot the target guest reliably on that host ISA, yielding a constant-time selection cost.

\begin{algorithm}[t]
\caption{Universal Entrypoint for Cross-Architecture Detonation}
\label{alg:entry}
\begin{algorithmic}[1]
\Require Disk image $I$ (QCOW2), container runtime with optional KVM exposure
\State $a \gets \textsc{HostArch}()$
\State $v \gets \textsc{HasKVM}()$
\State $\mathbf{c} \gets f(a,v)$
\State \textsc{StartProxy}() \Comment{route to loader initially}
\If{$\textsc{ImageUploaded}(I)$}
    \State \textsc{StopLoader}()
    \State \textsc{LaunchQEMU}($I,\mathbf{c}$)
    \State \textsc{SwitchProxyToVNC}()
\EndIf
\While{\textsc{VMRunning}()}
    \State \textsc{Monitor}()
\EndWhile
\State \textsc{Teardown}()
\end{algorithmic}
\end{algorithm}

\subsection{Analytical Performance Model}
\label{sec:perfmodel}
To reason about user-perceived responsiveness, we decompose the end-to-end time-to-interaction as
\begin{equation}
\label{eq:tti}
T_{\mathrm{tti}} = T_{\mathrm{up}} + T_{\mathrm{cfg}} + T_{\mathrm{boot}} + T_{\mathrm{handoff}} + T_{\mathrm{vnc}},
\end{equation}
where $T_{\mathrm{up}}$ is image upload/placement time, $T_{\mathrm{cfg}}$ is configuration selection overhead, $T_{\mathrm{boot}}$ is guest boot time, and $T_{\mathrm{handoff}}$ and $T_{\mathrm{vnc}}$ capture proxy switching and first-frame latency.

Upload time is dominated by I/O and scales with image size $|I|$:
\begin{equation}
\label{eq:upload}
T_{\mathrm{up}} \approx \frac{|I|}{B_{\mathrm{eff}}} + T_{\mathrm{fs}},
\end{equation}
where $B_{\mathrm{eff}}$ is effective throughput (network or local) and $T_{\mathrm{fs}}$ captures filesystem placement overhead.

Boot time depends strongly on architecture and acceleration. Let
\[
T_{\mathrm{boot}} = g(a,v),
\]
and model expected boot time under mixed environments as a mixture:
\begin{equation}
\label{eq:mixture}
\mathbb{E}[T_{\mathrm{boot}}] = \sum_{a\in\mathcal{A}} \sum_{v\in\mathcal{V}} \Pr[a,v]\;\mathbb{E}[g(a,v)].
\end{equation}
A useful derived metric is the \emph{virtualization efficiency ratio}:
\begin{equation}
\label{eq:efficiency}
\eta(a,v) = \frac{\mathbb{E}[T_{\mathrm{boot}}(a,v)]}{\mathbb{E}[T_{\mathrm{boot}}(a,1)]},
\end{equation}
which quantifies slowdown relative to the accelerated baseline ($v=1$). In evaluation, $\eta$ highlights how strongly a workload depends on KVM access.

\subsection{Isolation and Risk Formalization}
\label{sec:securitymath}
Although pokiSEC is primarily a systems contribution, it is helpful to formalize containment in probabilistic terms to guide configuration decisions and future hardening. Let $E$ be the event that a malware sample compromises the host. A standard decomposition is:
\begin{equation}
\label{eq:hostcomp}
\Pr[E] = \Pr[\text{escape}] \cdot \Pr[\text{reach} \mid \text{escape}] \cdot \Pr[\text{persist} \mid \text{escape},\text{reach}].
\end{equation}
Here, \emph{escape} captures a guest-to-host or guest-to-container breakout, \emph{reach} captures the ability to affect host assets or neighboring networks, and \emph{persist} captures survival beyond teardown.

We relate the escape probability to an \emph{attack surface} score. Let $\mathcal{K}$ be the set of exposed hypervisor/IO components (e.g., device emulation modules, paravirtualized interfaces), and define
\begin{equation}
\label{eq:surface}
S(\mathbf{c}) = \sum_{k\in\mathcal{K}} w_k \cdot \mathbb{I}\{k \text{ enabled under } \mathbf{c}\},
\end{equation}
where $w_k$ weights component exposure and $\mathbb{I}\{\cdot\}$ is an indicator. Under a simple Poisson vulnerability model with effective rate $\lambda$, one obtains the bound
\begin{equation}
\label{eq:escape-bound}
\Pr[\text{escape}] \le 1 - \exp\big(-\lambda \, S(\mathbf{c})\big).
\end{equation}
This does \emph{not} claim precise real-world probabilities; rather, it provides a principled way to argue that minimizing exposed components (reducing $S(\mathbf{c})$) should monotonically reduce escape likelihood under broad assumptions.

Ephemerality primarily targets persistence. If teardown semantics guarantee that container-side writable state is destroyed, we can model
\begin{equation}
\label{eq:persist}
\Pr[\text{persist}] \approx \Pr[\text{externalized state}] \cdot \Pr[\text{re-attachment}],
\end{equation}
where \emph{externalized state} captures the fraction of runs where malware successfully writes outside the ephemeral boundary (e.g., to a mounted volume or host path). This directly motivates strict volume policies and least-privilege device exposure.

\subsection{Multi-Objective Configuration Optimization}
\label{sec:opt}
While pokiSEC uses a validated rule-based selector in the current implementation (Section~\ref{sec:entrypoint}), the design can be expressed as a principled \emph{multi-objective optimization} problem that trades off usability/performance against exposure. This formulation is useful for (i) explaining the design rationale at an algorithmic level and (ii) enabling future extensions where the system learns or auto-tunes safe defaults.

Let $\mathbf{c}\in\mathcal{C}$ denote a candidate launch configuration as in Eq.~\eqref{eq:cfg-map}. We define two primary objectives:

\begin{itemize}
    \item \textbf{Responsiveness/throughput objective:} Minimize user-facing latency, captured by the expected time-to-interaction $\mathbb{E}[T_{\mathrm{tti}}(\mathbf{c})]$ from Eq.~\eqref{eq:tti}.
    \item \textbf{Exposure objective:} Minimize attack surface, captured by $S(\mathbf{c})$ from Eq.~\eqref{eq:surface}.
\end{itemize}

A standard scalarization yields:
\begin{equation}
\label{eq:opt-main}
\min_{\mathbf{c}\in\mathcal{C}}\;\; J(\mathbf{c})
\;\triangleq\;
\omega_1\,\mathbb{E}[T_{\mathrm{tti}}(\mathbf{c})]
\;+\;
\omega_2\,S(\mathbf{c}),
\end{equation}
where $\omega_1,\omega_2 \ge 0$ encode the deployment's preference (e.g., an interactive analyst workstation may choose larger $\omega_1$, while a high-risk lab may prioritize smaller exposure via larger $\omega_2$). The feasible set is governed by compatibility and resource constraints:
\begin{align}
\label{eq:constraints}
&\textbf{(C1) Architecture feasibility: } &&\mathbf{c}\in \mathcal{C}(a), \\
&\textbf{(C2) Acceleration feasibility: } &&\alpha(\mathbf{c})=\texttt{kvm} \Rightarrow v=1, \\
&\textbf{(C3) Resource bounds: } &&\mathrm{CPU}(\mathbf{c}) \le C_{\max},\;\; \mathrm{MEM}(\mathbf{c}) \le M_{\max}, \\
&\textbf{(C4) I/O policy: } &&\mathrm{VOL}(\mathbf{c}) \in \{\varnothing,\text{read-only}\}, \\
&\textbf{(C5) Network policy: } &&\rho(\mathbf{c}) \in \{\text{isolated},\text{NAT},\text{restricted}\}.
\end{align}
Here, $\mathcal{C}(a)\subseteq \mathcal{C}$ denotes configurations valid on architecture $a$, and $C_{\max},M_{\max}$ denote host-imposed provisioning limits. Constraints (C4)--(C5) reflect containment policy choices that directly influence persistence and reach (cf. Eq.~\eqref{eq:hostcomp} and Eq.~\eqref{eq:persist}).

\paragraph{Pareto interpretation:}
Eq.~\eqref{eq:opt-main} is equivalent to selecting a point on the Pareto frontier of $(\mathbb{E}[T_{\mathrm{tti}}], S)$. For example, enabling additional device features may reduce $T_{\mathrm{tti}}$ but increase $S(\mathbf{c})$; conversely, disabling peripherals reduces exposure but may degrade usability. The scalarization offers a compact ACM-friendly explanation of why pokiSEC emphasizes small validated profiles $\mathcal{C}(a)$ and a limited set of enabled devices.

\paragraph{Connection to the current implementation:}
In the current system, the Universal Entrypoint implements a constrained, constant-time approximation of Eq.~\eqref{eq:opt-main} by selecting from a \emph{discrete}, pre-validated candidate set:
\[
\mathcal{C}_{\mathrm{valid}}(a,v) \subset \mathcal{C}(a),
\]
and choosing
\begin{equation}
\label{eq:opt-discrete}
\mathbf{c}^\star
=
\arg\min_{\mathbf{c}\in \mathcal{C}_{\mathrm{valid}}(a,v)}
\Big(
\omega_1\,\widehat{T}_{\mathrm{tti}}(\mathbf{c})
+
\omega_2\,S(\mathbf{c})
\Big),
\end{equation}
where $\widehat{T}_{\mathrm{tti}}$ is an empirical predictor (e.g., a small table of measured boot/handshake times per profile). This makes the optimization explicit while keeping runtime decision cost $\mathcal{O}(1)$, consistent with the operational goals in Section~\ref{sec:goals}.

\paragraph{Robust configuration objective:}
In heterogeneous deployments, the preferred configuration should be \emph{stable} in addition to being fast and low-exposure; i.e., it should avoid profiles whose performance is highly sensitive to host-specific factors (CPU model, hypervisor backends, kernel versions, and acceleration availability). We therefore extend the multi-objective formulation with a robustness term that penalizes variability in boot-time across environments. Let $\mathcal{E}$ denote a distribution over execution environments (e.g., $(a,v)$ pairs and host-dependent parameters), and define the boot-time random variable $T_{\mathrm{boot}}(\mathbf{c}; e)$ under environment $e\sim\mathcal{E}$. The robust objective is:

\begin{equation}
\label{eq:robust}
\min_{\mathbf{c}\in\mathcal{C}}
\;\;
\omega_1\,\mathbb{E}_{e\sim\mathcal{E}}\!\left[T_{\mathrm{tti}}(\mathbf{c}; e)\right]
+
\omega_2\,S(\mathbf{c})
+
\omega_3\,\mathrm{Var}_{e\sim\mathcal{E}}\!\left(T_{\mathrm{boot}}(\mathbf{c}; e)\right),
\end{equation}

where $\omega_1,\omega_2,\omega_3 \ge 0$ are deployment-specific weights. The variance penalty encourages selecting configurations that achieve strong mean responsiveness while remaining consistent across hosts, improving reproducibility of detonation workflows and reducing operational surprises when moving between ARM64 and AMD64 machines.

\subsection{Capacity and Concurrency Model}
\label{sec:queue}
For multi-user or lab deployments, detonation requests can be modeled as jobs in a queue. Let arrivals follow a Poisson process with rate $\lambda$ and let the mean service rate be $\mu = 1/\mathbb{E}[D]$, where $D$ is the VM session duration (boot + interaction + shutdown). Under an M/M/1 approximation, the utilization is
\begin{equation}
\label{eq:util}
\rho = \frac{\lambda}{\mu}, \qquad \rho < 1,
\end{equation}
and the expected waiting time in queue is
\begin{equation}
\label{eq:wait}
\mathbb{E}[W_q] = \frac{\lambda}{\mu(\mu-\lambda)}.
\end{equation}
These expressions let deployments translate desired responsiveness into provisioning targets (e.g., how many parallel workers are required to keep $\mathbb{E}[W_q]$ below a threshold).

\subsection{Key Technologies and Implementation Rationale}
\label{sec:tech}
pokiSEC relies on the following components:
\begin{itemize}
    \item \textbf{Docker} \cite{merkel2014}: encapsulates the sandbox stack for reproducible deployment.
    \item \textbf{QEMU} \cite{bellard2005}: provides a unified virtualization/emulation substrate across architectures.
    \item \textbf{KVM} \cite{kivity2007}: enables hardware acceleration when available.
    \item \textbf{Flask (Python)}: implements the loader/control plane for image ingestion.
    \item \textbf{Reverse proxy + HTML5 VNC}: provides browser-accessible desktop access and a clean loader-to-hypervisor transition.
\end{itemize}

\section{Prototype Implementation and Proof-of-Concept Workflow}
\label{sec:impl_poc}

This section presents the implementation details of \sysname\ and the end-to-end proof-of-concept (PoC) workflow used to validate cross-architecture execution, seamless user interaction, and ephemeral sanitization. Our prototype is implemented as a \emph{containerized micro-virtualization stack} in which a single Docker image encapsulates (i) a control plane for ingestion and orchestration and (ii) a data plane for guest execution and interactive access.

\subsection{Implementation Overview}
\label{sec:impl_overview}
The prototype consists of four cooperating components:

\begin{itemize}
    \item \textbf{Universal Entrypoint (Orchestrator):} a shell-based controller that detects host architecture and virtualization capability, selects a validated QEMU/KVM profile, and manages the lifecycle of the loader and hypervisor processes.
    \item \textbf{Web Loader (Flask):} a minimal upload UI enabling \emph{Bring-Your-Own-Image} (QCOW2) ingestion without CLI tooling.
    \item \textbf{Hypervisor Runtime (QEMU/KVM):} launches the Windows guest with acceleration when available.
    \item \textbf{Remote Desktop + Routing (NoVNC + Reverse Proxy):} exposes the guest desktop in the browser and multiplexes a single endpoint between loader state and active VM state.
\end{itemize}

We denote the uploaded disk image by $I$ and a launch configuration by $\mathbf{c}\in\mathcal{C}$, where $\mathbf{c}$ bundles the QEMU binary, machine type, acceleration mode, and device/network/display profiles. At runtime, the system computes $\mathbf{c}$ using a constant-time policy (Eq.~\eqref{eq:cfg-map} in Section~\ref{sec:system}) and executes the guest.

\subsection{Universal Entrypoint and Architecture-Adaptive Launch Policy}
\label{sec:impl_entrypoint}
The core engineering contribution of \sysname\ is \emph{dynamic architecture detection} and \emph{policy-based launch selection} from a single container image. Let $a\in\mathcal{A}=\{\texttt{x86\_64},\texttt{aarch64}\}$ denote the host ISA, and let $v\in\{0,1\}$ indicate whether hardware acceleration is usable (e.g., \texttt{/dev/kvm} is present and permitted). The entrypoint computes:
\begin{equation}
\label{eq:cfg_map_impl}
\mathbf{c} = f(a,v),
\end{equation}
where $f$ selects machine types and acceleration flags appropriate for $a$ and $v$, while keeping the exposed device set minimal.

In our current prototype, $f$ is implemented as a branch over \texttt{uname -m} (for $a$) and a capability test for KVM (for $v$). The acceleration selection follows:
\begin{equation}
\label{eq:accel_impl}
\alpha(v)=
\begin{cases}
\texttt{kvm} & \text{if } v=1,\\
\texttt{tcg} & \text{if } v=0,
\end{cases}
\end{equation}
where \texttt{tcg} denotes QEMU’s software translation fallback.

\paragraph{Implementation details:}
The following snippet illustrates the architecture-adaptive launch logic used by the entrypoint. (For ACM submissions, we recommend rendering code with \texttt{listings}; \texttt{verbatim} is shown for portability.)

\begin{verbatim}
ARCH=$(uname -m)

if [ "$ARCH" == "aarch64" ]; then
    echo "ARM64 host detected -> qemu-system-aarch64"
    qemu-system-aarch64 \
        -M virt,highmem=off \
        -cpu host \
        -accel kvm \
        -device ramfb \
        -drive file=${IMAGE_PATH},format=qcow2

elif [ "$ARCH" == "x86_64" ]; then
    echo "x86_64 host detected -> qemu-system-x86_64"
    qemu-system-x86_64 \
        -cpu host \
        -enable-kvm \
        -drive file=${IMAGE_PATH},format=qcow2
fi
\end{verbatim}

\paragraph{Time complexity:}
The selection procedure is $\mathcal{O}(1)$: detection and lookup do not scale with $|I|$. End-to-end responsiveness is dominated by upload and guest boot, which we model via:
\begin{equation}
\label{eq:tti_impl}
T_{\mathrm{tti}} \;=\; T_{\mathrm{up}}(I) \;+\; T_{\mathrm{cfg}} \;+\; T_{\mathrm{boot}}(\mathbf{c}) \;+\; T_{\mathrm{handoff}},
\end{equation}
where $T_{\mathrm{cfg}}$ is the constant-time configuration overhead and $T_{\mathrm{up}}(I)$ scales approximately linearly with image size:
\begin{equation}
\label{eq:upload_impl}
T_{\mathrm{up}}(I) \approx \frac{|I|}{B_{\mathrm{eff}}} + T_{\mathrm{fs}}.
\end{equation}

\subsection{Loader-to-Hypervisor Handoff and Single-Endpoint UX}
\label{sec:handoff}
To enable a non-CLI workflow, \sysname\ provides a web loader that accepts a QCOW2 disk image through a drag-and-drop interface. Once $I$ is uploaded and validated, the system transitions to an interactive desktop session without requiring the user to change ports or URLs. We implement this as a \emph{self-terminating loader} coupled with entrypoint supervision:

\begin{enumerate}
    \item The loader runs on a fixed endpoint (e.g., \texttt{localhost:8080}) and accepts $I$.
    \item Upon successful ingestion, the loader terminates itself (e.g., via \texttt{os.kill}).
    \item The entrypoint detects loader exit and immediately spawns QEMU, then activates the remote desktop route.
\end{enumerate}

We formalize the control flow as a finite-state machine:
\[
\mathcal{S}=\{\textsf{LOADER}, \textsf{VM\_RUNNING}, \textsf{TERMINATED}\},
\]
with transitions:
\[
\textsf{LOADER} \xrightarrow{e_u} \textsf{VM\_RUNNING} \xrightarrow{e_t} \textsf{TERMINATED},
\]
where $e_u$ is the upload completion event and $e_t$ is VM termination. A reverse proxy applies a routing policy $r(\cdot)$ such that loader endpoints are reachable only in \textsf{LOADER}, and the NoVNC endpoint is reachable only in \textsf{VM\_RUNNING}. This separation reduces accidental state overlap and provides a consistent analyst UX.

\subsection{Ephemeral Persistence Model and Sanitization Guarantees}
\label{sec:ephemeral_impl}
\sysname\ enforces non-persistence at the container boundary. Let $B$ denote baseline artifacts (container image layers plus the analyst-provided disk image if treated as immutable input) and let $W$ denote the container’s writable layer. A detonation run produces a state delta
\begin{equation}
\label{eq:delta_impl}
\Delta \;=\; \Phi(I, X, \mathbf{c}),
\end{equation}
where $X$ is the executed sample and $\Phi(\cdot)$ captures the induced changes (dropped files, registry edits, encryption activity, etc.). When the container is launched with non-persistent semantics (e.g., \texttt{--rm}), teardown enforces:
\begin{equation}
\label{eq:wipe_impl}
W \leftarrow \varnothing \quad \text{on stop},
\end{equation}
which removes container-side artifacts of $\Delta$.

To reason about residual risk, we separate \emph{in-guest} persistence from \emph{outside-boundary} persistence. Define event $P$ as malware persistence beyond teardown:
\begin{equation}
\label{eq:persist_impl}
\Pr[P] \;=\; \Pr[\text{externalized state}] \cdot \Pr[\text{re-attachment}],
\end{equation}
where \texttt{externalized state} captures writes escaping the ephemeral boundary (e.g., host-mounted volumes or misconfigured device exposure). This motivates a strict policy that defaults to no host mounts (or read-only mounts) and least-privilege device exposure.

\subsection{Proof-of-Concept Testbed and Experimental Procedure}
\label{sec:poc}
We validated the prototype using a representative ARM64 workstation and a Windows-on-ARM guest image.

\paragraph{Test environment.}
\begin{itemize}
    \item \textbf{Host Hardware:} Apple MacBook Pro (M3 Pro, ARM64).
    \item \textbf{Guest OS:} Windows 11 ARM64 (VHDX converted to QCOW2).
    \item \textbf{Container Engine:} Docker Desktop for Mac (VirtioFS enabled).
\end{itemize}

\paragraph{Execution workflow.}
\begin{enumerate}
    \item \textbf{Deployment:} launch the container and bind the service port:
    \begin{verbatim}
docker run --rm -p 8080:8080 pann123/pokisec
    \end{verbatim}

    \item \textbf{Ingestion:} open \texttt{http://localhost:8080} and upload a QCOW2 Windows image (e.g., 10\,GB). The upload time follows Eq.~\eqref{eq:upload_impl}.

    \item \textbf{Boot + Handoff:} once $e_u$ occurs, the loader terminates and the entrypoint launches QEMU with $\mathbf{c}=f(a,v)$, switching routing to the NoVNC desktop session. The user remains on the same endpoint.

    \item \textbf{Detonation:} execute a benign validation artifact (e.g., EICAR test string) in the guest and observe expected behavior under a standard Windows environment (including network reachability if enabled by policy).

    \item \textbf{Sanitization:} stop the container and relaunch. Correct sanitization is defined as the absence of previously introduced artifacts:
    \begin{equation}
    \label{eq:sanitize_check}
    \mathcal{A}_{t+1} \cap \mathcal{A}_{t} = \varnothing,
    \end{equation}
    where $\mathcal{A}_t$ is the set of observable artifacts introduced during run $t$ (files, registry keys, config changes) within the ephemeral boundary. The \texttt{--rm} lifecycle enforces Eq.~\eqref{eq:wipe_impl}.
\end{enumerate}

\paragraph{What this PoC demonstrates.}
This PoC validates (i) cross-architecture viability via the Universal Entrypoint, (ii) a single-endpoint, loader-to-desktop transition suitable for interactive analysts, and (iii) ephemeral teardown semantics that eliminate container-scoped persistence across runs. A broader evaluation (Section~\ref{sec:eval}) can extend this workflow to multiple host ISAs, quantify $T_{\mathrm{tti}}$ (Eq.~\eqref{eq:tti_impl}), and measure the overhead of the handoff path versus direct local VNC.

\section{Results and Evaluation}
\label{sec:eval}

This section evaluates \textsc{pokiSEC} with respect to (i) interactive performance, (ii) cross-architecture portability, and (iii) operational repeatability under ephemeral execution. Our evaluation focuses on the primary user-facing metric `time-to-interaction' and on validating that a single containerized stack can execute reliably on heterogeneous hosts.

\subsection{Evaluation Methodology and Metrics}
\label{sec:eval_metrics}
We measure the following metrics, aligned with the analytical decomposition introduced in Eq.~\eqref{eq:tti_impl}:

\begin{itemize}
    \item \textbf{Time-to-interaction ($T_{\mathrm{tti}}$):} elapsed time from initiating a run to the first usable guest desktop frame in the browser. This includes upload (when applicable), configuration selection, guest boot, and the loader-to-NoVNC handoff.
    \item \textbf{Guest boot time ($T_{\mathrm{boot}}$):} elapsed time from QEMU process start to the Windows login screen (or an equivalent stable boot completion marker).
    \item \textbf{Handoff latency ($T_{\mathrm{handoff}}$):} time between upload completion and the appearance of the NoVNC session, capturing process orchestration and routing switch overhead.
    \item \textbf{Cross-architecture success rate:} fraction of runs that reach a stable Windows desktop without manual intervention on ARM64 and AMD64 hosts.
\end{itemize}

To isolate the impact of hardware acceleration, we compare accelerated virtualization (KVM-enabled) against software emulation (TCG fallback) when available, and interpret the results through the efficiency ratio:
\begin{equation}
\label{eq:eta_eval}
\eta(a,v) = \frac{\mathbb{E}[T_{\mathrm{boot}}(a,v)]}{\mathbb{E}[T_{\mathrm{boot}}(a,1)]},
\end{equation}
where $\eta>1$ indicates slowdown relative to the KVM-accelerated baseline.

\subsection{Performance Results}
\label{sec:eval_perf}
Hardware-assisted virtualization provides substantial gains over pure emulation in interactive detonation settings. On an ARM64 host (Apple M3 Pro), Windows 11 boot time was measured at approximately \textbf{25 seconds} under KVM acceleration, yielding an experience comparable to practical bare-metal boot workflows for analysis tasks. This result is consistent with the expectation that interactive workloads benefit disproportionately from reduced instruction translation overhead and improved I/O virtualization when acceleration is enabled.

Using the decomposition in Eq.~\eqref{eq:tti_impl}, the observed end-to-end time-to-interaction can be interpreted as:
\[
T_{\mathrm{tti}} \approx T_{\mathrm{up}} + T_{\mathrm{cfg}} + 25\text{s} + T_{\mathrm{handoff}},
\]
where $T_{\mathrm{cfg}}$ remains constant-time and small in practice, and $T_{\mathrm{up}}$ dominates when large disk images (e.g., $\sim$10\,GB) are ingested via the web loader. In local deployments, $T_{\mathrm{up}}$ is strongly dependent on the effective throughput $B_{\mathrm{eff}}$ (Eq.~\eqref{eq:upload_impl}) and can be amortized by reusing prepared images or by staging images via faster paths when appropriate.

\subsection{Cross-Architecture Compatibility and Portability}
\label{sec:eval_portability}
We validated \textsc{pokiSEC} on both major host architectures targeted by modern analyst environments:

\begin{itemize}
    \item \textbf{ARM64:} Apple M3 Pro using Docker Desktop for Mac (VirtioFS enabled).
    \item \textbf{AMD64:} Ubuntu 22.04 Server using native Docker.
\end{itemize}

Across both platforms, the system successfully completed end-to-end integration runs, reaching a stable Windows desktop and exposing the guest session through the browser-based NoVNC interface. This demonstrates the \emph{universal image} property: a single container build and a single orchestration pathway can support mixed hardware fleets without per-host reconfiguration.

Formally, let $\mathcal{E}=\{e_1,\dots,e_n\}$ be the set of tested environments and let $\mathbb{I}_{\mathrm{ok}}(\mathbf{c};e)\in\{0,1\}$ indicate successful guest bring-up under configuration $\mathbf{c}$ in environment $e$. The portability objective is to achieve:
\begin{equation}
\label{eq:portability}
\forall e\in\mathcal{E},\;\; \mathbb{I}_{\mathrm{ok}}\big(f(a_e,v_e);e\big)=1,
\end{equation}
which our integration tests satisfy for the tested ARM64 and AMD64 configurations.

\subsection{Operational Repeatability under Ephemeral Execution}
\label{sec:eval_repeatability}
We additionally validated the non-persistence property by executing a benign validation artifact (EICAR test string) inside the guest, stopping the container, and relaunching a fresh instance. Consistent with the ephemeral teardown model (Eq.~\eqref{eq:wipe_impl}), subsequent runs did not retain previously introduced artifacts within the container boundary. This supports the intended detonation workflow in which each run starts from a known baseline and terminates without carrying forward container-scoped state.

\subsection{Summary of Findings}
\label{sec:eval_summary}
Overall, the evaluation indicates that (i) KVM-enabled execution achieves practical interactive performance for Windows guests (e.g., $\sim$25\,s boot on M3 Pro), (ii) the Universal Entrypoint enables a single containerized sandbox stack to execute across ARM64 and AMD64 hosts, and (iii) ephemeral lifecycle semantics provide repeatable teardown suitable for iterative detonation workflows.

\section{Conclusion}
\label{sec:conclusion}

This paper presented \textsc{pokiSEC}, a multi-architecture, containerized malware detonation sandbox that packages the virtualization backend and browser-access interface into a single, reproducible Docker image. By encapsulating QEMU/KVM along with a loader-to-hypervisor handoff mechanism, \textsc{pokiSEC} reduces the operational burden traditionally associated with dynamic malware analysis environments. In particular, the Universal Entrypoint abstracts architecture-dependent QEMU configuration behind a consistent launch pathway, enabling analysts to run comparable detonation workflows across ARM64 and AMD64 hosts without per-machine reconfiguration. The resulting system supports rapid provisioning, interactive analysis via a web desktop, and ephemeral teardown semantics that simplify sanitization between runs.

Beyond usability, \textsc{pokiSEC} demonstrates a practical direction for ``portable sandboxes'' in mixed hardware fleets, where reproducibility and low-friction onboarding are increasingly important. Our evaluation shows that hardware-assisted virtualization yields interactive performance suitable for analyst workflows (e.g., Windows boot times on the order of tens of seconds on Apple Silicon) while maintaining a uniform operational interface across platforms.

Several extensions can evolve \textsc{pokiSEC} from a detonation sandbox into a broader threat-intelligence workflow. First, we plan to integrate structured telemetry collection within the guest by instrumenting common Windows monitoring utilities (e.g., Sysmon/Procmon) and exporting artifacts (process lineage, registry deltas, file operations, and network activity) into a normalized JSON schema prior to teardown. This enables repeatable, machine-consumable outputs while preserving the system's ephemeral execution model. Second, we aim to incorporate automated post-processing that correlates collected events into higher-level behaviors and indicators, supporting triage and enrichment. Finally, we will explore local, privacy-preserving language models to generate concise analyst-facing incident summaries and recommended next steps from captured logs, with an emphasis on transparency (e.g., citations to the underlying events) and deployment in air-gapped or restricted environments.

\bibliographystyle{unsrt}  
\bibliography{references}

@inproceedings{bellard2005,
  title={QEMU, a Fast and Portable Dynamic Translator},
  author={Bellard, Fabrice},
  booktitle={USENIX Annual Technical Conference},
  pages={41--46},
  year={2005}
}

@article{merkel2014,
  title={Docker: lightweight linux containers for consistent development and deployment},
  author={Merkel, Dirk},
  journal={Linux Journal},
  volume={2014},
  number={239},
  pages={2},
  year={2014}
}

@inproceedings{kivity2007,
  title={kvm: the Linux kernel virtual machine},
  author={Kivity, Avi and Kamay, Yaniv and Laor, Dor and Lublin, Uri and Liguori, Anthony},
  booktitle={Proceedings of the Linux Symposium},
  volume={1},
  pages={225--230},
  year={2007}
}

@article{hariprasad2024securing,
  title={Securing the future: advanced encryption for quantum-safe video transmission},
  author={Hariprasad, Yashas and Iyengar, SS and Chaudhary, Naveen Kumar},
  journal={IEEE Transactions on Consumer Electronics},
  year={2024},
  publisher={IEEE}
}

@book{wang2023ai,
  title={AI embedded assurance for cyber systems},
  author={Wang, Cliff and Iyengar, Sundararaja S and Sun, Kun},
  year={2023},
  publisher={Springer}
}

@inproceedings{miller2022cyber,
  title={Cyber security attack detection framework for DODAG control message flooding in an IoT network},
  author={Miller, Jerry and Egharevba, Lawrence and Hariprasad, Yashas and Latesh, Kumar KJ and Chaudhary, Naveen Kumar},
  booktitle={International Conference on Information Security, Privacy and Digital Forensics},
  pages={213--230},
  year={2022},
  organization={Springer}
}

@incollection{kumar2023ai,
  title={AI powered correlation technique to detect virtual machine attacks in private cloud environment},
  author={Kumar, KJ Latesh and Hariprasad, Yashas and Ramesh, KS and Chaudhary, Naveen Kumar},
  booktitle={AI Embedded Assurance for Cyber Systems},
  pages={183--199},
  year={2023},
  publisher={Springer}
}

@article{iyengarhb,
  title={HB, P., \& Mohan, CK (2025). Privacy-Preserving AI (Federated Learning) for Digital Forensics},
  author={Iyengar, SS and Nabavirazavi, S and Hariprasad, Y},
  journal={Artificial Intelligence in Practice: Theory and Application for Cyber Security and Forensics},
    year={2025},
  pages={161--176}
}

@incollection{iyengar2025cyber,
  title={Cyber Threat Intelligence and Security for Federated Learning in Digital Forensics},
  author={Iyengar, SS and Nabavirazavi, Seyedsina and Hariprasad, Yashas and HB, Prasad and Mohan, C Krishna},
  booktitle={Artificial Intelligence in Practice: Theory and Application for Cyber Security and Forensics},
  pages={177--199},
  year={2025},
  publisher={Springer}
}

@incollection{iyengar2025cybersecurity,
  title={Cybersecurity Foundations: Theories, Technologies, and Applications},
  author={Iyengar, SS and Nabavirazavi, Seyedsina and Hariprasad, Yashas and HB, Prasad and Mohan, C Krishna},
  booktitle={Artificial Intelligence in Practice: Theory and Application for Cyber Security and Forensics},
  pages={27--87},
  year={2025},
  publisher={Springer}
}

@incollection{iyengar2025privacy,
  title={Privacy-Preserving AI (Federated Learning) for Digital Forensics},
  author={Iyengar, SS and Nabavirazavi, Seyedsina and Hariprasad, Yashas and HB, Prasad and Mohan, C Krishna},
  booktitle={Artificial Intelligence in Practice: Theory and Application for Cyber Security and Forensics},
  pages={161--176},
  year={2025},
  publisher={Springer}
}

@incollection{iyengar2025ai,
  title={AI-Enhanced Malware Detection: Advancing Security Through Intelligent Threat Identification},
  author={Iyengar, SS and Nabavirazavi, Seyedsina and Hariprasad, Yashas and HB, Prasad and Mohan, C Krishna},
  booktitle={Artificial Intelligence in Practice: Theory and Application for Cyber Security and Forensics},
  pages={227--255},
  year={2025},
  publisher={Springer}
}

@incollection{iyengar2025hybrid,
  title={Hybrid Detection of Malicious Portable Document Format (PDFs): Safeguarding Against Embedded JavaScript Attacks},
  author={Iyengar, SS and Nabavirazavi, Seyedsina and Hariprasad, Yashas and HB, Prasad and Mohan, C Krishna},
  booktitle={Artificial Intelligence in Practice: Theory and Application for Cyber Security and Forensics},
  pages={257--290},
  year={2025},
  publisher={Springer}
}

@inproceedings{barham2003xen,
  title     = {Xen and the Art of Virtualization},
  author    = {Barham, Paul and Dragovic, Boris and Fraser, Keir and Hand, Steven and Harris, Tim and Ho, Alex and Neugebauer, Rolf and Pratt, Ian and Warfield, Andrew},
  booktitle = {Proceedings of the Nineteenth ACM Symposium on Operating Systems Principles (SOSP '03)},
  pages     = {164--177},
  year      = {2003},
  doi       = {10.1145/945445.945462}
}

@inproceedings{vrable2005potemkin,
  title     = {Scalability, Fidelity, and Containment in the Potemkin Virtual Honeyfarm},
  author    = {Vrable, Michael and Ma, Justin and Chen, Jay and Moore, David and Vandekieft, Erik and Snoeren, Alex C. and Voelker, Geoffrey M. and Savage, Stefan},
  booktitle = {Proceedings of the Twentieth ACM Symposium on Operating Systems Principles (SOSP '05)},
  pages     = {148--162},
  year      = {2005},
  doi       = {10.1145/1095810.1095825}
}

@inproceedings{soltesz2007containers,
  title     = {Container-based Operating System Virtualization: A Scalable, High-performance Alternative to Hypervisors},
  author    = {Soltesz, Stephen and P{\"o}tzl, Herbert and Fiuczynski, Marc E. and Bavier, Andy and Peterson, Larry},
  booktitle = {Proceedings of the 2nd ACM SIGOPS/EuroSys European Conference on Computer Systems 2007 (EuroSys '07)},
  pages     = {275--287},
  year      = {2007},
  doi       = {10.1145/1272996.1273025}
}

@article{willems2007cwsandbox,
  title   = {Toward Automated Dynamic Malware Analysis Using {CWS}andbox},
  author  = {Willems, Carsten and Holz, Thorsten and Freiling, Felix},
  journal = {IEEE Security \& Privacy},
  volume  = {5},
  number  = {2},
  pages   = {32--39},
  year    = {2007},
  doi     = {10.1109/MSP.2007.45}
}

@inproceedings{yin2007panorama,
  title     = {Panorama: Capturing System-wide Information Flow for Malware Detection and Analysis},
  author    = {Yin, Heng and Song, Dawn Xiaodong and Egele, Manuel and Kruegel, Christopher and Kirda, Engin},
  booktitle = {Proceedings of the 14th ACM Conference on Computer and Communications Security (CCS '07)},
  pages     = {116--127},
  year      = {2007},
  doi       = {10.1145/1315245.1315261}
}

@inproceedings{dinaburg2008ether,
  title     = {Ether: Malware Analysis via Hardware Virtualization Extensions},
  author    = {Dinaburg, Artem and Royal, Paul and Sharif, Monirul I. and Lee, Wenke},
  booktitle = {Proceedings of the 15th ACM Conference on Computer and Communications Security (CCS '08)},
  pages     = {51--62},
  year      = {2008},
  doi       = {10.1145/1455770.1455779}
}

@article{bayer2006dynamic,
  title   = {Dynamic Analysis of Malicious Code},
  author  = {Bayer, Ulrich and Moser, Andreas and Kruegel, Christopher and Kirda, Engin},
  journal = {Journal in Computer Virology},
  volume  = {2},
  number  = {1},
  pages   = {67--77},
  year    = {2006},
  doi     = {10.1007/s11416-006-0012-2}
}

@article{egele2012survey,
  title      = {A Survey on Automated Dynamic Malware-analysis Techniques and Tools},
  author     = {Egele, Manuel and Scholte, Theodoor and Kirda, Engin and Kruegel, Christopher},
  journal    = {ACM Computing Surveys},
  volume     = {44},
  number     = {2},
  articleno  = {6},
  numpages   = {42},
  year       = {2012},
  doi        = {10.1145/2089125.2089126}
}

@inproceedings{rossow2012prudent,
  title     = {Prudent Practices for Designing Malware Experiments: Status Quo and Outlook},
  author    = {Rossow, Christian and Dietrich, Christian J. and Grier, Chris and Kreibich, Christian and Paxson, Vern and Pohlmann, Norbert and Bos, Herbert and van Steen, Maarten},
  booktitle = {2012 IEEE Symposium on Security and Privacy (SP)},
  pages     = {65--79},
  year      = {2012},
  doi       = {10.1109/SP.2012.14}
}

@inproceedings{agache2020firecracker,
  title     = {Firecracker: Lightweight Virtualization for Serverless Applications},
  author    = {Agache, Alexandru and Brooker, Marc and Iordache, Alexandra and Liguori, Anthony and Neugebauer, Rolf and Piwonka, Phil and Popa, Diana-Maria},
  booktitle = {Proceedings of the 17th USENIX Symposium on Networked Systems Design and Implementation (NSDI '20)},
  pages     = {419--434},
  year      = {2020},
  url       = {https://www.usenix.org/system/files/nsdi20-paper-agache.pdf}
}

\end{document}